\documentclass[12pt]{article}

\usepackage[margin=0.7in]{geometry}
\usepackage[utf8]{inputenc}
\usepackage{amsmath}
\usepackage{amsfonts}
\usepackage{bm}
\usepackage{mathtools}
\usepackage{graphicx}
\usepackage[authoryear]{natbib}

\usepackage{import}
\usepackage{standalone}
\usepackage{tabularx}
\usepackage{lscape}
\usepackage{longtable}
\usepackage{array,multirow}
\usepackage{authblk}
\usepackage{tcolorbox}
\usepackage{pgfplots}
\usepackage{tikzscale}
\usepackage{pdfpages}
\usepackage{makecell}
\usepackage{mdframed}
\usepackage{enumitem}%
\usepackage{dsfont}
\usepackage{amsthm}
\usepackage{stmaryrd}
\usepackage{amssymb}
\usepackage{hyperref}

\usepackage{graphicx}
\usepackage{tcolorbox}
\usepackage{tikzscale}
\usetikzlibrary{arrows, automata}

\usepackage{soul}
\usepackage{xcolor}

\usepackage{parskip}

\newcommand{\type}{\mathsf{type}}
\newcommand{\samp}{\mathsf{samp}}

\newcommand{\outlier}{\mathsf{outlier}}
\newcommand{\source}{\mathsf{source}}
\newcommand{\charic}{\mathsf{char}}
\newcommand{\PMA}{\mathsf{PMA}}

\mathtoolsset{showonlyrefs}

\begin{document}

\title{Temporal Models for Demographic and Global Health Outcomes in Multiple Populations: Introducing the Normal-with-Optional-Shrinkage Data Model Class}

\author[1]{Leontine Alkema\thanks{This work was supported, in whole or in part, by the Bill \& Melinda Gates Foundation (INV-00844). Under the grant conditions of the Foundation, a Creative Commons Attribution 4.0 Generic License has already been assigned to the Author Accepted Manuscript version that might arise from this submission. 
Contact: lalkema@umass.edu.}}
\affil[1]{\textit{Department of Biostatistics and Epidemiology, School of Public Health and Health Sciences, University of Massachusetts Amherst, Amherst, Massachusetts, USA}}

\author[2] {Herbert Susmann}
\affil[2]{\textit{Division of Biostatistics, Department of Population Health, NYU Grossman School of Medicine, New York, New York, USA
}}

\author[1]{Evan Ray}

\date{
November 26, 2024
}

\maketitle

\abstract{
Statistical models are used to produce estimates of demographic and global health indicators in populations with limited data. Such models integrate multiple data sources to produce estimates and forecasts with uncertainty based on model assumptions. Model assumptions can be divided into assumptions that describe latent trends in the indicator of interest versus assumptions on the data generating process of the observed data, conditional on the latent process value. Focusing on the latter, we introduce a class of data models that can be used to combine data from multiple sources with various reporting issues. The proposed data model accounts for sampling errors and differences in observational uncertainty based on survey characteristics. In addition, the data model employs horseshoe priors to produce estimates that are robust to outlying observations. We refer to the data model class as the normal-with-optional-shrinkage (NOS) set up.  We illustrate the use of the NOS data model for the estimation of modern contraceptive use and other family planning indicators at the national level for countries globally, using survey data. 
}

\section{Introduction}
Measuring and tracking demographic and health indicators is of wide interest. One  major application lies in assessing progress towards meeting international goals (such as the Sustainable Development Goals) and identifying areas where improvement is needed. Population-level demographic and health related data may come from a number of sources, including vital registration systems, health records, and surveys. Combining data from multiple data sources in order to produce integrated estimates and forecasts of indicators of interest is a significant challenge for which statistical models are often used. 

Many statistical models for estimating and projecting demographic and health indicators exhibit a common structure, in which it is possible to decompose a model into two parts: a \textit{process model} and a \textit{data model} \citep{susmann2021temporal}. The process model describes how the true, latent value of an indicator is expected to evolve over time. For example, the process model might impose an assumption that a demographic indicator will evolve relatively smoothly over time. As the true value of the indicator is unobserved, the role of the data model is to describe how the observations (e.g. from surveys) relate to the true value. The data model may incorporate assumptions about the bias and variance of the observations. If observations are derived from multiple sources, the data model may have additional structure to handle specific prior knowledge about each data source. Once the general framework of considering a model as a combination of a process and data model is established, there are still a myriad of choices to be made to fully specify the structure of each component. 

In this paper, we focus on specification of the data model. We introduce a flexible class of data models referred to as \textit{normal-with-optional-shrinkage} (NOS) models that can be used to integrate data from multiple sources that may exhibit various types of reporting issues. One major type of issue that must be accounted for is uncertainty in the observations in the form of measurement variance. The NOS model can account for various types of measurement variance, including for example design-based measurement error from complex survey designs. In addition, the data model explicitly accounts for extreme outlying observations, which increases the robustness of the final model estimates. We illustrate the use of the NOS data model for the estimation of family planning indicators at the national level across many countries, using multiple sources of survey data. We show in detail how the NOS model can be designed to handle various data issues that arise in this context.

The rest of the paper is organized as follows. The following section motivates the need for data models by introducing an indicator, modern contraceptive use, the data sources that are used to form estimates of this indicator, and the data issues that need to be accounted for in this specific context. The following methods section reviews the overall modeling framework and defines the NOS data model class. We continue with a case study in which we develop a specific NOS model that we use to combine multiple family planning data sources to produce estimates of country-level modern contraceptive use and other family planning indicators. We conclude with a discussion of the data model, case study, and future work.

\section{Motivating example}
To motivate the need for statistical models that can handle multiple data sources of various quality, we consider as an example a family planning (FP) indicator that is of wide interest. We return to this indicator later in the paper when we build a model for estimating and projecting a set of FP indicators in countries.

The motivating indicator is the proportion of married or in-union women aged 15-49 within a population who are using (or whose partner is using) a modern contraceptive method. For brevity, this indicator is referred to as \textit{mCPR}. Modern methods of contraception include female and male sterilization, oral hormonal pills, the intra-uterine device, male and female condoms, injectables, the implant, vaginal barrier methods, standard days method, lactational amenorrhea method, and emergency contraception. 

Data on mCPR may be obtained from household surveys conducted by international organizations or by local organizations or governments. Data on mCPR is obtained in such surveys  based on questions in which participants report whether they or their partner are currently using a modern contraceptive method. Further information on survey data in provided in \cite{alkema_2024evol}. In summary, survey programs can be categorized into the Demographic and Health Survey Program (DHS), Performance Monitoring for Action (PMA), UNICEF Multiple Indicator Cluster Surveys (MICS), national surveys, and other surveys. Micro data from each survey can be used to calculate estimates and sampling errors of mCPR for the population-time period covered by the survey, taking into account the complex sampling design of each survey. Measurement errors in mCPR survey estimates may arise from multiple sources, e.g., based on how the survey data were collected or compiled, or if the population sampled by the survey differs from the target population of interest. 

\begin{figure}[htbp]
    \centering
    \includegraphics[width=0.44\textwidth]{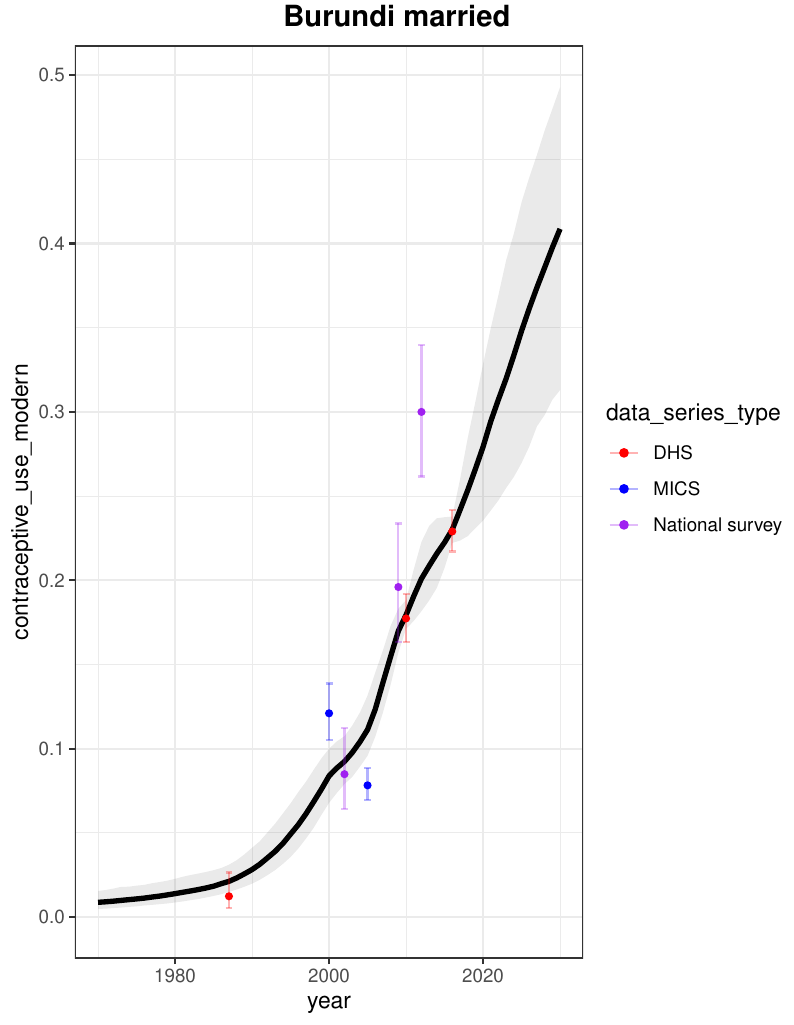}
    \includegraphics[width=0.44\textwidth]{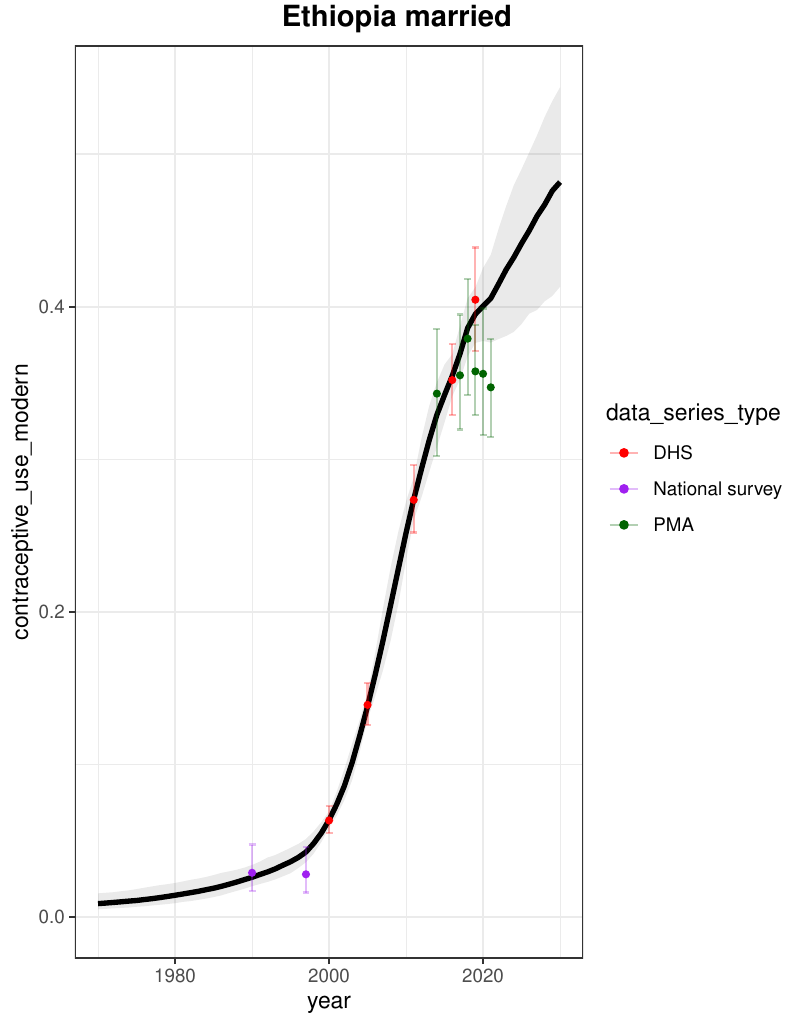}
    \caption{\textbf{Data and estimates for mCPR among married women for Burundi and Ethiopia.} Data are shown in colored dots with 95\% confidence intervals that reflect sampling errors. FPET estimates are shown in black, with grey shaded areas representing 90\% credible intervals. 
    }
    \label{fig-data}
\end{figure}

Estimates from single surveys may not be sufficient to inform programs and understand progress; rather, it is typically necessary to aggregate estimates from multiple surveys in order to have a broader understanding of trends. Examples are shown for mCPR in Figure~\ref{fig-data} for Burundi and Ethiopia, showing how results from different surveys may not be in complete agreement. The figure shows estimates of mCPR derived from multiple surveys. In Burundi, the most recent national survey suggests a large fluctuation in mCPR which is likely to be due to measurement error. For Ethiopia, we see that the most recent DHS provides an estimate of mCPR around 41\% for 2019 (95\% confidence interval based on sampling errors alone is given by (37, 44)), while the most recent PMA survey states that mCPR is around 35\% (32, 38) in 2021. 
The sampling errors for Ethiopia are not that small so confidence intervals  overlap but only barely so, calling into question whether one of the two sources is off for some other reason.  

The Burundi and Ethiopia examples illustrate that data sources need to be combined to produce well-founded estimates and forecasts with uncertainty that can be used to inform programs and understand progress towards targets. There may be various reasons why combining data from multiple sources is advantageous, including high uncertainty/measurement error associated with individual data sources or lack of recent data. In such settings, models can be used to combine information from different data sources and to produce estimates and predictions of the true latent value of the indicator, along with uncertainty intervals.

Estimates and predictions of mCPR have been produced for various populations using Bayesian modeling techniques. Initial work was focused on a global assessment of mCPR using survey data \citep{alkema2013mcpr}. For in-country usage in low- and middle-income countries, the Family Planning Estimation Tool (FPET) was developed, which is a country-specific implementation of a global model for FP estimation \citep{new_levels_2017}. Subsequently, the global model and FPET evolved to better capture local contexts \citep{cahill2018mcpr, kantorova_estimating_2020}. FPET fits, produced by its most recent version \citep{alkema_2024evol}, are shown in Figure~\ref{fig-data}. Based on FPET assumptions, the fit in Burundi smooths over the outlying national survey. In Ethiopia, the FPET fit aligns with the most recent DHS and trends above the PMA data series. FPET assumptions are based on a normal-with-optional-shrinkage (NOS) data model, introduced in the next section. 

\section{Methods}

\subsection{Problem statement}
We start by introducing notation to formalize the problem of producing estimates and forecasts of time-varying indicators in multiple populations. Let the indicator of interest be denoted by $\phi_{c,t}$ for population (e.g., country) $c$ at time $t$, with $c = 1, \dots, C$ and $t = 1, \dots, T$. In the motivating example, $\phi_{c,t}$ refers to the true mCPR in country $c$ and year $t$. Observations are denoted $y_i$, $i = 1, \dots, N$. We assume that each observation is associated with a single population and time point, denoted $c[i]$ and $t[i]$ for the population (country) and time point of observation $i$, respectively. Each observation typically has additional metadata associated with it, such as the type of survey it was derived from. Going back to the motivating example to illustrate, $y_i$ refers to the mCPR measured by the $i$th survey, and the metadata includes its survey type (DHS, PMA, ...).

\begin{figure}[htbp]
    \centering
    \includegraphics[width=\textwidth, height=0.55\textwidth]{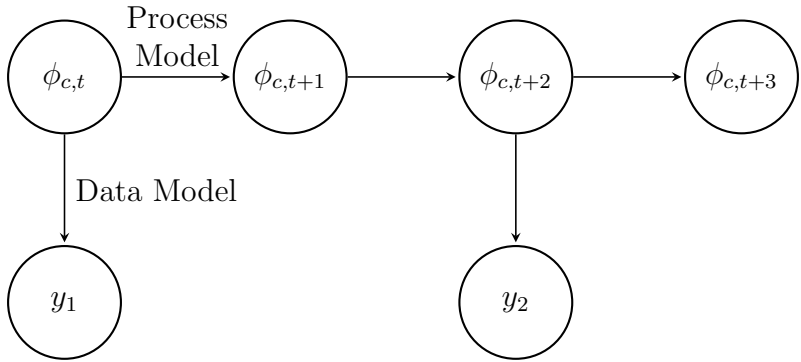}
    \caption{Figure adapted from \citep{susmann2021temporal}. The TMMP model class distinguishes between the true values of an indicator $\phi_{c,t}$ and the noisy observed data $y_i$. The process model describes the evolution of the true values, and the data model describes how the observed data are generated from the true values. This structure handles missing data naturally: the latent trend is modeled for all time points by the process model, and it is possible that only some time points have observed values. In this example, observed data only exist for times $t$ and $t+2$. \label{fig-tmmp}} 
\end{figure}

Estimates and forecasts of the indicator of interest are derived by inferring the parameters $\phi_{c,t}$ conditional on the observations $y_i$. In order to do so, we must impose a model on the $\phi_{c,t}$ and the relationship between the $\phi_{c,t}$ and the observations $y_i$. Following the Temporal Models for Multiple Populations (TMMPs; \citealt{susmann2021temporal}) framework, we refer to these two components of the model as \textit{process model} and a \textit{data model}  \citep{susmann2021temporal}.
The process model defines the evolution of $\phi_{c,t}$ over time and across populations and the data model encodes the relationship between $\phi_{c,t}$ and observations $y_i$. A conceptual overview is shown in Figure~\ref{fig-tmmp}.

Various options for the process model were previously considered in \cite{susmann2021temporal}. In this work, we focus on the data model. In the next section, we describe a general class of data models that may incorporate data from multiple sources exhibiting various data reporting issues.

\subsection{Normal-with-optional-shrinkage (NOS) data model class}
In this section we introduce a general class of data models, which we refer to as \textit{normal-with-optional-shrinkage} (NOS) models. The NOS class is designed to incorporate data from multiple sources and that may have various types of reporting error.

To illustrate the various types of measurement error that may arise, consider the case of surveys. Survey data are typically subject to multiple sources of error. First, all survey estimates are subject to sampling error arising from the survey design. 
In addition, other non-systematic errors, referred to as measurement errors, may occur. Measurement errors arise due to errors introduced during data collection or compilation. In some cases, this error may be extreme, leading to estimates of the indicator that are severely out-of-trend with estimates from other surveys. Error may also occur if the sampled population differs from the population of interest. 

We define observation error as the difference between the observed value and the underlying latent true value of the indicator of interest, possibly on a transformed scale, such that the range of the errors is not constrained. We subsequently decompose the error to be the sum of errors arising from multiple error types. By default, we assume independent normal distributions for each individual error type. For errors arising from the presence of outliers we use a modified approach based on shrinkage priors. Conditional dependence between total errors across observations may introduced where needed, for example when observations are collected as part of a longitudinal data collection scheme and observation errors may be correlated over time.  

Formally, let $E_i$ be the total (unobserved) error corresponding to observation $i$, defined using transformation function $h(\cdot)$:
\begin{align}
    E_i &= h(y_i) - h(\phi_{c[i], t[i]}).
\end{align}
For example, a logit-transform could be applied for indicators constrained to fall in $(0, 1)$ or a log-transform for positive indicators.

Let $\mathcal{T}$ be a set containing the relevant types of relevant observation error. For example, $\mathcal{T}$ may include design-based sampling error and survey measurement errors. We refer generically to a specific type of error as $\type \in \mathcal{T}$. Let $\mathcal{I}^{(\type)}$ be the set of observations that are deemed subject to $\type$. We then decompose the total error $E_i$ as a sum of the error of each type:
\begin{align}
    E_i & = \sum_{\type \in \mathcal{T}} E_i^{(\type)} \mathbb{I}(i  \in \mathcal{I}^{(\type)}),
\end{align}
where $\mathbb{I}(\cdot)$ is the indicator function. This decomposition of the total error is very flexible in that it allows for arbitrary types of error, and each error type may apply to a different set of observations. 

In the NOS data model, each type-specific error term is assumed to be normally distributed with mean zero and standard deviation $\sigma_i^{(\type)}$: 
\begin{align}
    E_i^{(\type)}|\sigma_i^{(\type)}  &\sim N\left(0, \sigma_i^{(\type)2}\right). 
\end{align}
That the normal distributions are centered at zero implies an assumption that the observations are unbiased on the transformed scale. This assumption can be weakened as needed, for example to allow for systematically biased observations. We assume that observation-type-specific errors are conditionally independent, such that the resulting conditional density for the total error (and therefore the transformed observations) is also normal:
\begin{equation}
h(y_i)|h(\phi_{c[i],t[i]}), \sigma_i \sim N(h(\phi_{c[i], t[i]}), \sigma_i^2), \label{dm-normal} 
\end{equation}
with $\sigma_i^2 = Var(E_i|\sigma_i^{(\type)}) = \sum_{\type \in \mathcal{T}} \sigma_i^{(\type)2} \mathbb{I}(i  \in \mathcal{I}^{(\type)})$.
If errors are correlated across observations, the following multivariate data model is used:
\begin{equation}
    h(\bm{y})|h(\bm{\phi}), \bm{\Sigma} \sim N(h(\bm{\phi}), \bm{\Sigma}), \label{dm-mvn}
\end{equation}
where $h(\bm{\phi})$ refers to the respective mean vector of $h(\bm{y})$, and $\bm{\Sigma}$ to its associated covariance matrix with $\Sigma_{j,l} = \sigma_j \cdot \sigma_l \cdot \rho_{j,l}$. Specification of correlation $\rho_{j,l}$ depends on the application; an example is provided for estimating mCPR in the case study below.

We distinguish between three types of error variances: (i) sampling, (ii) repeated, and (iii) outlier error variances. We take each one in turn below.

\paragraph{Sampling errors} Sampling variance refers to variance arising from the survey design, and is calculated based on survey micro data. Sampling variance is a fixed value, and is not inferred as a model parameter. We denote sampling variance as $\sigma_i^{(\samp)2} = s_i^2$, where $s_i^2$ is the pre-computed survey sampling variance. 

\paragraph{Repeated errors} Repeated errors have a variance that is shared across observations. These types of error may be based on survey characteristics, such as the particular survey program that the observation belongs to. For mCPR estimation, for example, the survey program MICS has been been found to have higher variance than DHS \citep{alkema2013mcpr}. In such settings, the error variances are unknown data model parameters that are shared across subsets of observations. Formally, we denote these variances as $\sigma_i^{(\type)} =\tilde{\sigma}_{d[i]}^{(\type)}$ if observation $i$ has survey characteristic $d[i]$.

\paragraph{Outlier errors} Outlier error variance is an unknown observation-specific error variance for capturing outlier observations. In brief, the outlier error variance arises from a shrinkage prior on the error terms that shrinks outlier errors towards zero for most observations, only allowing large errors when the observation is strongly different from the trend. In this set up, the standard deviation term, denoted by $\sigma_i^{(\outlier)}$, is parameterized using a scale parameter $\tau$ that is shared across observations, and an observation-specific scale parameter $\gamma_i$ that allows individual errors to escape from shrinkage. The set up is explained in detail below.

\paragraph{Details on outlier errors}
Some observations may be subject to extreme reporting errors, which may arise if for example some part of the survey design or implementation is poorly executed. A data model assuming normally distributed errors based on sampling and source-type-specific variance terms alone is not well suited to deal with outliers that are extreme relative to the total standard deviation associated with the normal error. The effect of using  normally distributed error terms in settings where outliers are present is that estimates are pulled towards the outlying data points, as opposed to smoothing over them. 

To allow for possible outliers we use an approach inspired by regularization in regression modeling. The idea is to use densities that encourage shrinkage of outlier error terms to zero for most observations, but allow large errors for observations that are extreme outliers. The expression for the  distribution of an outlier error $E_i^{(\outlier)}$ follows that of a regularized horseshoe prior, as specified by \cite{piironen2017horseshoe}:
\begin{align}
E_i^{(\outlier)} \mid \gamma_{i}, \tau, \vartheta &\sim N(0, \tau^{2} \tilde{\gamma}_{i}^{2}), \\
\tilde{\gamma}_{i}^{2} &= \frac{\vartheta^{2} \gamma_{i}^{2}}{\vartheta^{2} + \tau^2 \gamma_{i}^{2}}, \end{align}
such that 
\begin{align}
    \sigma_i^{(\outlier)}= \tau \tilde{\gamma}_{i} = \sqrt{\frac{\tau^2 \vartheta^{2} \gamma_{i}^{2}}{\vartheta^{2} + \tau^2 \gamma_{i}^{2}}}.
\end{align}
In the above expression, the parameter $\tau^{}$ is a \textit{global scale} parameter which shrinks all of the outlier error terms towards zero.  Individual outlier errors may escape this shrinkage through the \textit{local scale} parameter $\gamma^{}_{i}$, which is assigned a heavy-tailed half-Cauchy prior:
\begin{align}
\gamma_{i}^{} &\sim C^+(0, 1).
\end{align}
The transformation $\tilde{\gamma}^{}_{i}$ of $\gamma_{i}^{}$ allows for regularization of the values of outlier errors that escape global shrinkage, in that for large $\gamma_{i}^{2}$, the prior on the errors approaches a normal prior with variance $\vartheta^{2}$, i.e.
\begin{align}
    \sigma^{(\outlier)2}_i \to \vartheta^{2} \text{ as } \gamma_i \to \infty,\\
   \sigma^{(\outlier)2}_i \to 0 \text{ as } \gamma_i \to 0
\end{align}
The choice of the set $\mathcal{I}^{(\outlier)}$, which refers to the set of data points that may subject to extreme errors, should be guided by the application. We discuss the set up for estimating mCPR in the next section.

\section{Case study}
As a case study we present a specific instance of an NOS data model designed to handle multiple data sources for family planning indicators. Specifically, we seek to estimate the proportion of married women who use a modern method of contraception (mCPR), as well as several additional related family planning (FP) indicators. These indicators categorize women into (1) women who use a modern method, (2) women who have an unmet need for modern contraceptives, and (3) women who do not have a demand for family planning (referred to as the no need group). For further information on the definition of these categories, see \cite{alkema_2024evol}. 

Formally, denote the respective proportions in each family planning category as $\phi_j$ for $j=1,2,3$ referring to modern, unmet need, and no need categories, respectively, and note that $\sum_j \phi_j=1$.

To formalize the observed data, let the vector of observed FP proportions be denoted by $y_{i,j}$ where $y_{i,1}$ refers to mCPR, $y_{i,2}$ to unmet need for modern methods, and finally, $y_{i,3}$ to women without a demand for family planning.
The survey source of each observation is denoted $d[i] \in \{ 1, \dots, D \}$, i.e., DHS, MICS, PMA, National survey, or other survey.

To produce estimates and predictions for these FP indicators, we build a statistical model that is the combination of a process model and a data model. An overview of the process model, which encodes assumptions about how $\phi_j$ evolves over time in each population, can be found in \cite{alkema_2024evol}. Our main focus is in showing how an NOS data model can effectively handle the types of data reporting issues that arise in this case study.

\subsection{Data model}
We apply an NOS data model to model the (logit-transformed) observed mCPR $y_{i,1}$. A separate NOS data model is used to model the ratio of (logit-transformed) unmet need for modern methods over non-modern-users $y_{i,2}/(1-y_{i, 1})$. 

With slight misuse of notation for ease of exposition, let $y_i$ refer to one observation of an indicator of interest in the set of FP indicators defined above, e.g., observed mCPR or the ratio of unmet need to non-modern users, and let $\phi_{c[i], t[i]}$ refer to the latent value of the indicator for the corresponding country-year. Define the transformed observation
\begin{align}
z_i &= \mathrm{logit}(y_i),\\
& = \mathrm{logit}\left(\phi_{c[i], t[i]}\right) + E_i,
\end{align}
where $E_i$ is the total error on the logit-scale. We decompose $E_i$ into four types of error:
\begin{equation}\label{dm-fp}
    E_i = E_i^{(\samp)} + \mathbb{I}(d[i] \in \mathcal{D})E_{i}^{(\source)} + \mathbb{I}(i \in \mathcal{C})E_{i}^{(\charic)} + \mathbb{I}(i \in \mathcal{O})E_i^{(\outlier)}, 
\end{equation}
where sampling error is given by $E_i^{(\samp)}$, source-type specific measurement error by $E_{i}^{(\source)}$, $E_{i}^{(\charic)}$ captures differences if the sampled population differs from the population of interest, and finally, observation-specific error $E_i^{(\outlier)}$ captures outliers. Indicator functions indicate whether the error term is assigned to each observation; details on the index sets $\mathcal{D}$, $\mathcal{C}$, and $\mathcal{O}$ are provided below.

For the sampling error $E_i^{(\samp)}$, its associated variance $\sigma_i^{(\samp)2} = s_i^2$, where $s_i$ is the sampling error on the logit-scale. The logit-scale sampling error is derived from the sampling error for the observed proportion using the delta method. Where possible, survey micro data are processed to obtain sampling errors. If micro data are not available, imputation is based on estimates of effective samples size, i.e., sample size and design effects. 

Source-type errors are designed to capture cases where variance from some survey programs may be larger than others:
\begin{align}
    \sigma_i^{(\source)} = \tilde{\sigma}_{d[i]}^{(\source)}.
\end{align}
For FP estimation, we assign source errors to observations from all sources except those from DHS, i.e., $\mathcal{D}$ excludes DHS. The following categories were considered: MICS, PMA, national surveys, and other survey programs. 

For FP estimation, some surveys do not capture the exact population of interest. For example, this may occur if ever-married women were interviewed rather than currently married women, or if age groups differ (see \cite{alkema2013mcpr,kantorova_estimating_2020}). In those instances, an additional term $E_i^{(\charic)}$ is added to capture this difference, with variance $\sigma^{(\charic)2}$. The subset of observation indices for which the population surveyed differs from that of interest, $\mathcal{C}$, is obtained from survey metadata.  

The final  error term is designed to handle detection of outlier observations. As a preprocessing step, we first categorize observations into two categories: ``possibly outlying'' or ``not possibly outlying''. In the data model in Eq.~\eqref{dm-fp}, observation indices $\mathcal{O}$ refer to the set of data points that may be possibly outlying, the set of indices of observations that have an outlier error term assigned. This categorization is introduced to, at a country level, avoid smoothing over high-quality observations that are deemed to capture true fluctuations in FP indicators, and, at a global level, avoid global error parameters being influenced by high-quality data that are not representative of the targeted category of possibly outlying observations. Details of this preprocessing step are given in the appendix. The outlier error term is assigned a horseshoe shrinkage prior as described previously.

Given the expression for the error terms in Eq.~\eqref{dm-fp}, and assuming independence across errors for a single observation, the density for individual observations is as in Eq.~\eqref{dm-normal}. We assume observations are conditionally independent except for data collected by the PMA program.  Analysis of PMA data has suggested that PMA data series may differ consistently from other survey data, and that results from consecutive PMA surveys may be correlated due to repeated sampling of individuals from the same clusters across different survey rounds \footnote{PMA sampling methodology is described at \url{https://www.pmadata.org/data/survey-methodology}.}. %
Figure 1 shows one possible example of this effect, in Ethiopia. For PMA data series for a country-marital group, we use the multivariate normal specification:
\begin{align}
    \bm{z}|\bm{\phi}, \bm{\Sigma} \sim N(\mathrm{logit}(\bm{\phi}), \bm{\Sigma}), 
\end{align}
where $\mathrm{logit}(\bm{\phi})$ refers to the respective mean vector of $\bm{z}$, and $\bm{\Sigma}$ to its associated covariance matrix. We set $\Sigma_{j,l} = Cov(z_j, z_l) = \sigma_j \cdot \sigma_l \cdot \rho_{\PMA}^{|t[j] - t[l]|}$, where  $\rho_{\PMA}$ refers to the autocorrelation in PMA data errors, and $\sigma_j$ combines all of the contributions to the marginal variance of $z_j$ from the error decomposition described above.

\paragraph{Priors}
To complete the Bayesian model specification, priors on the data model parameters are as follows:
\begin{align}
   \sigma_{d}^{(source)}  &\sim N^+(0, 0.5^2),\\
   \tau &\sim C^+(0, 0.04^2),\\
   \vartheta &\sim N^+(0,1),\\
   \rho_{PMA} &\sim U(0,1).   
\end{align}

\paragraph{Computation}
The model was implemented in the \texttt{Stan} programming language \citep{stan2023, cmdstanr2022}, and all analyses were conducted using the \texttt{R} statistical computing environment \citep{r2022}. Further details are provided in \cite{alkema_2024evol}.

\subsection{Results}
\subsubsection{Global findings related to data quality for estimating mCPR}          

\begin{figure}[htbp]
    \centering
    \includegraphics[width=0.6\textwidth]{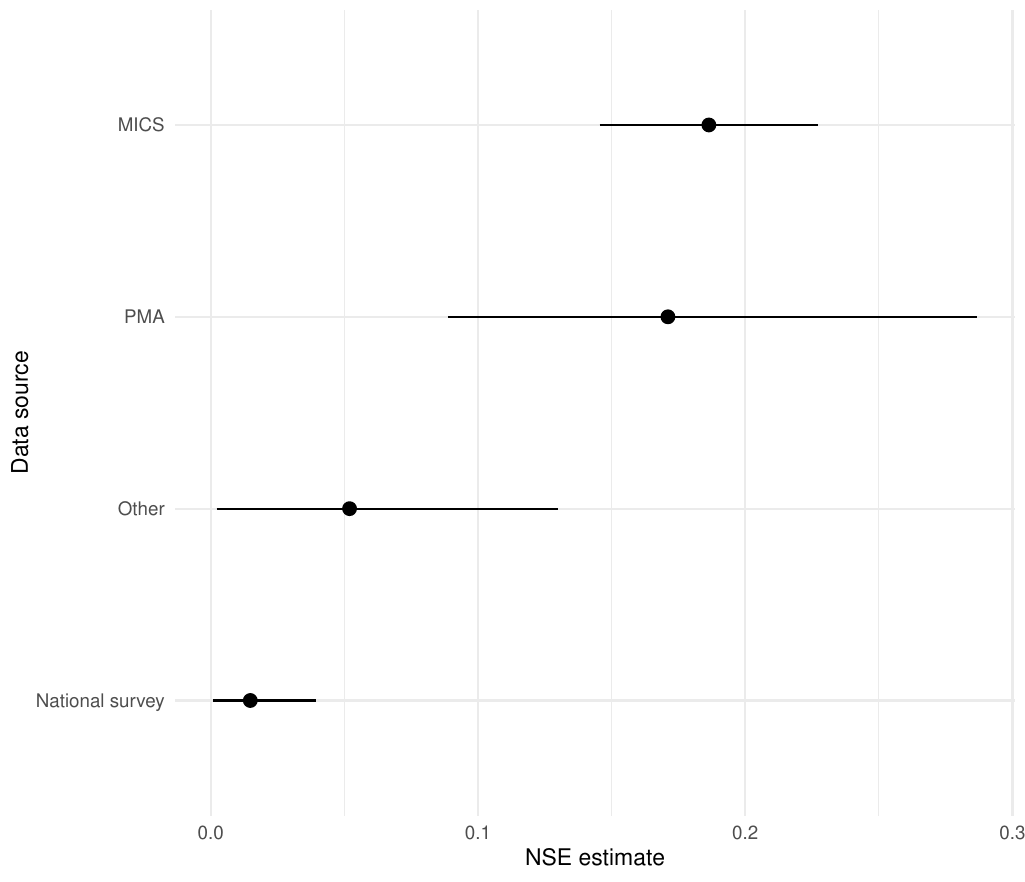}
    \caption{\textbf{Estimates of source-specific standard deviation terms}. Mean estimate with 95\% credible intervals.}
    \label{fig-source}
\end{figure}

Estimates of source-type variance parameters for estimating mCPR are given in Figure~\ref{fig-source}. The estimates range from 0.015 (95\% CI given by [0, 0.04]) for national surveys to 0.19 [0.15, 0.23] for observations from MICS. Correlation in PMA data series is substantial: the correlation parameter $\rho_{PMA}$ is estimated at 0.80 [0.57, 0.94]. 

\begin{figure}[htbp]
    \centering
    \includegraphics[width=0.8\textwidth]{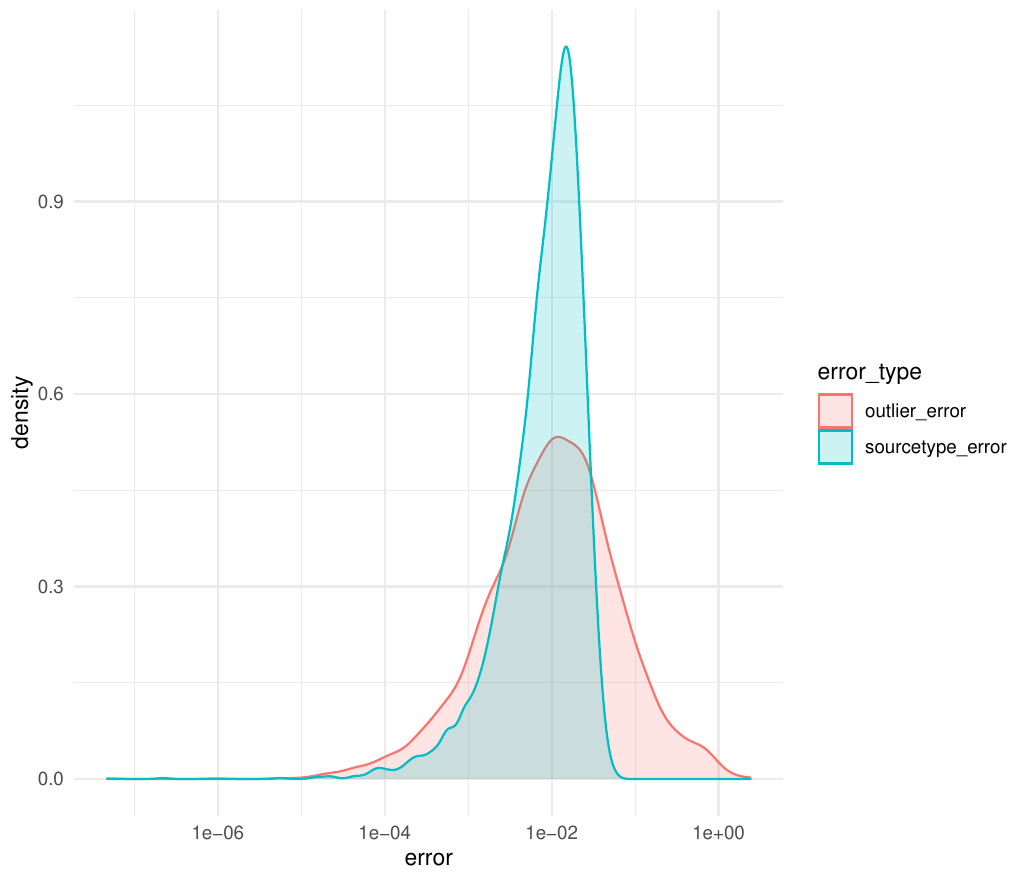}
    \caption{\textbf{Predictive densities for outlier errors and source-type errors for observations from national surveys.} Density refers to log10(absolute errors), with the horizontal axis labeled in units of error.}
    \label{fig-outliers}
\end{figure}

Figure~\ref{fig-outliers} visualizes the predictive density for outlier errors and errors based on source errors from national surveys. Error densities are shown for absolute errors (referring to the absolute value of the error on the logit-scale), plotted on the log10-scale for legibility. As per the data model design, the predictive density for outlier errors shows higher probability of large errors.

\subsubsection{Illustrative country estimates and forecasts}    
Estimates are shown for the all countries in Appendix Figure~\ref{fig-fits}. In this section, we show illustrative examples to illustrate how data availability and quality affects the fits. Uncertainty associated with the observations shown in the figures is based on total error variance, i.e., the total uncertainty associated with the observation.

\begin{figure}[htbp]
    \centering
    \includegraphics[width=0.6\textwidth]{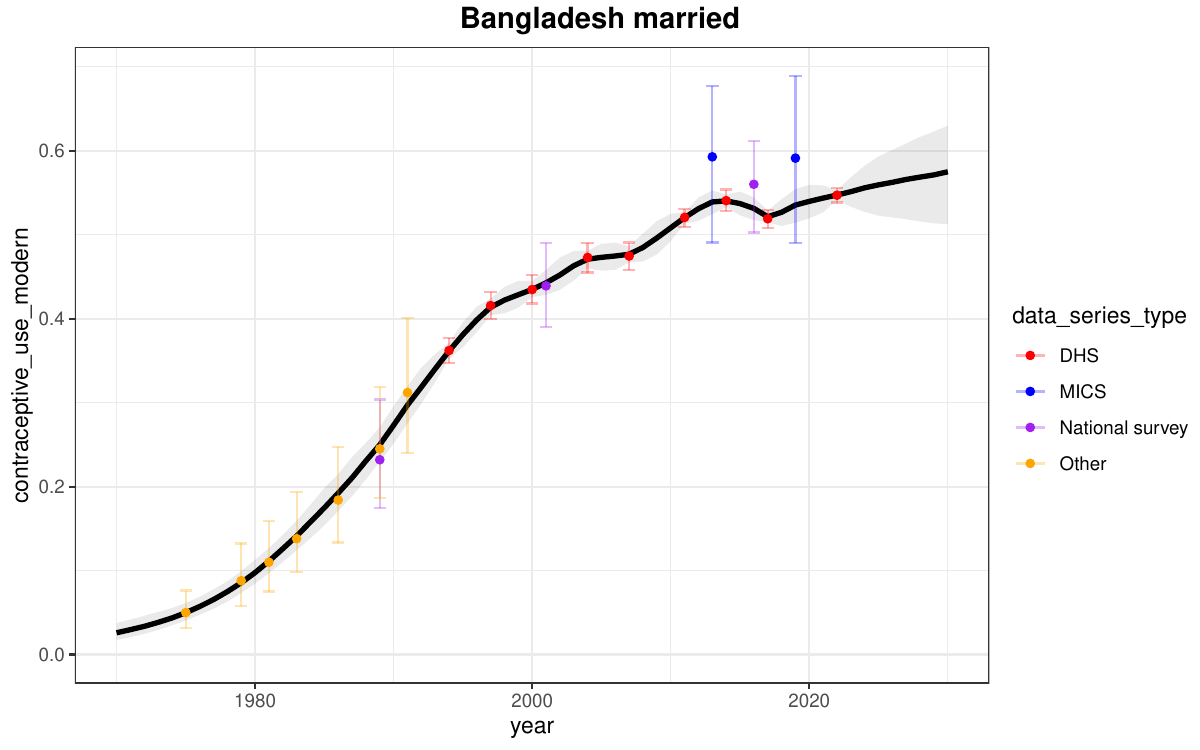}\\
    \includegraphics[width=0.6\textwidth]{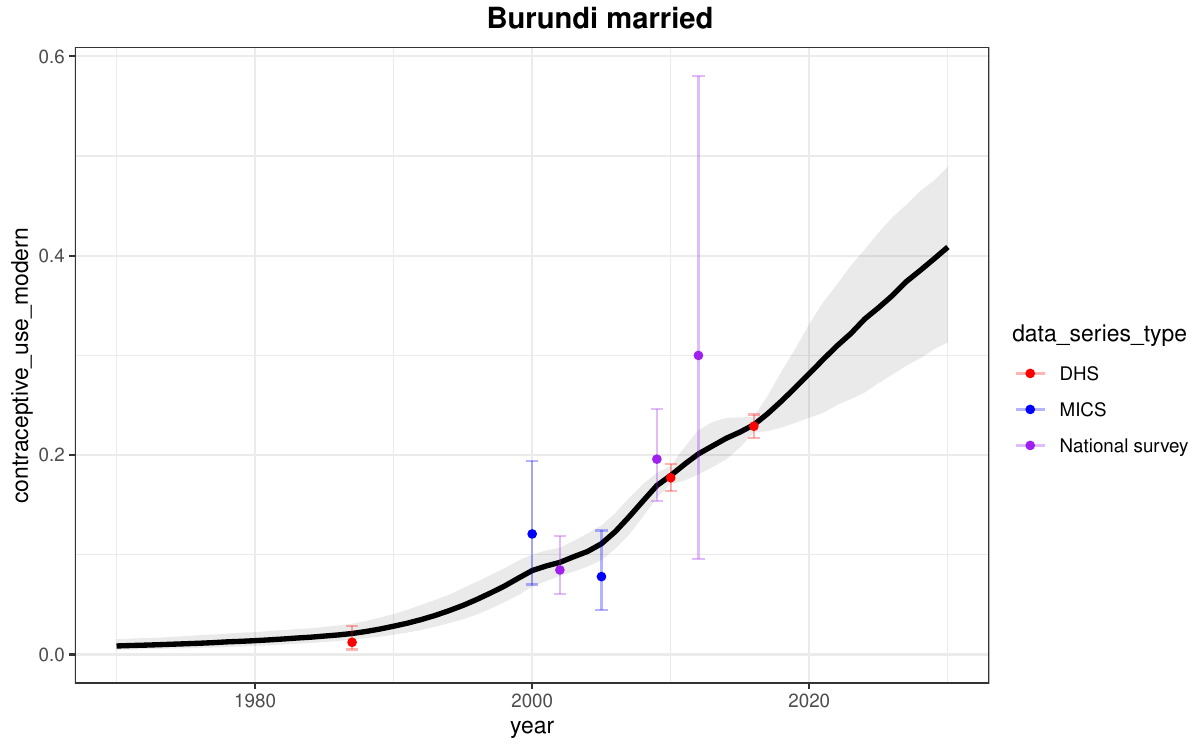}\\
     \includegraphics[width=0.6\textwidth]{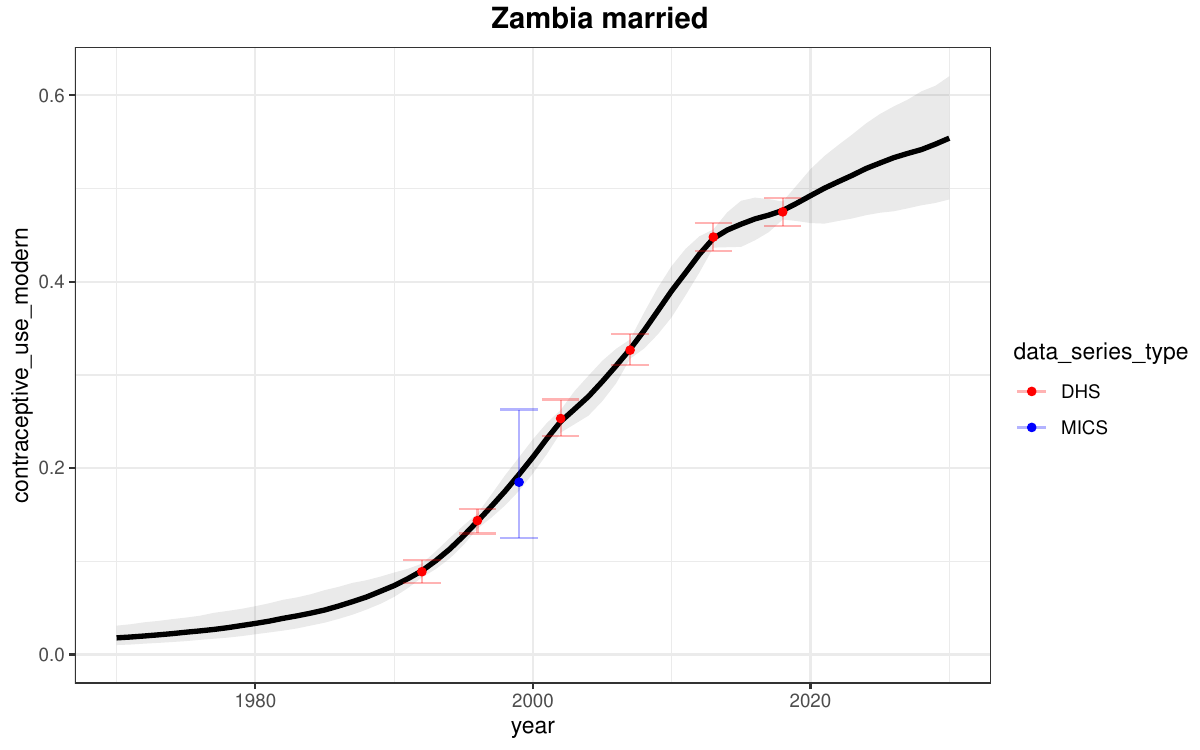}
   \caption{\textbf{Data and estimates for mCPR among married women for Bangladesh, Burundi, 
and Zambia.} Data are shown in colored dots with 95\% confidence intervals that reflect total error variance. FPET estimates are shown in black, with grey shaded areas representing 90\% credible intervals. In the selected countries, estimates are primarily informed by DHS data.}
    \label{fig-fits-dhs}
\end{figure}

In Figure~\ref{fig-fits-dhs}, fits are shown for Bangladesh, Burundi, 
and Zambia. In these countries, estimates are primarily informed by DHS data. Burundi is an example where a national survey is outlying; the model estimates smooth over this data point. 

\begin{figure}[htbp]
    \centering
    \includegraphics[width=0.45\textwidth]{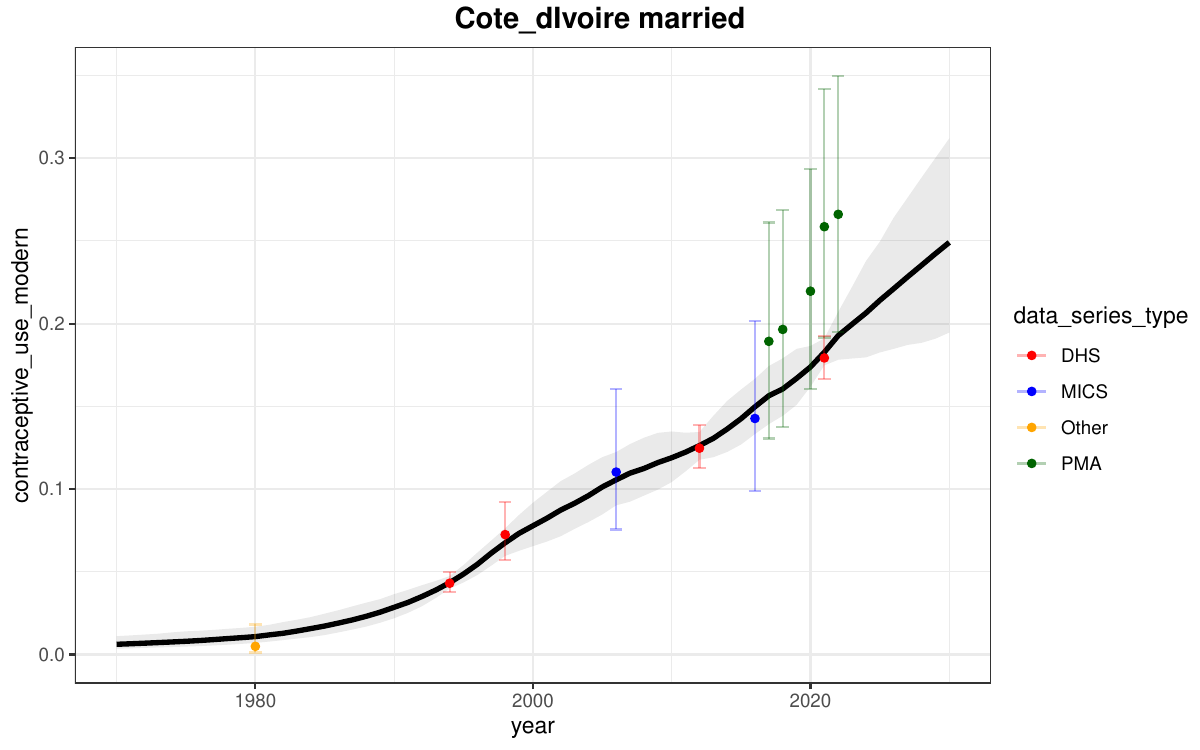} 
    \includegraphics[width=0.45\textwidth]{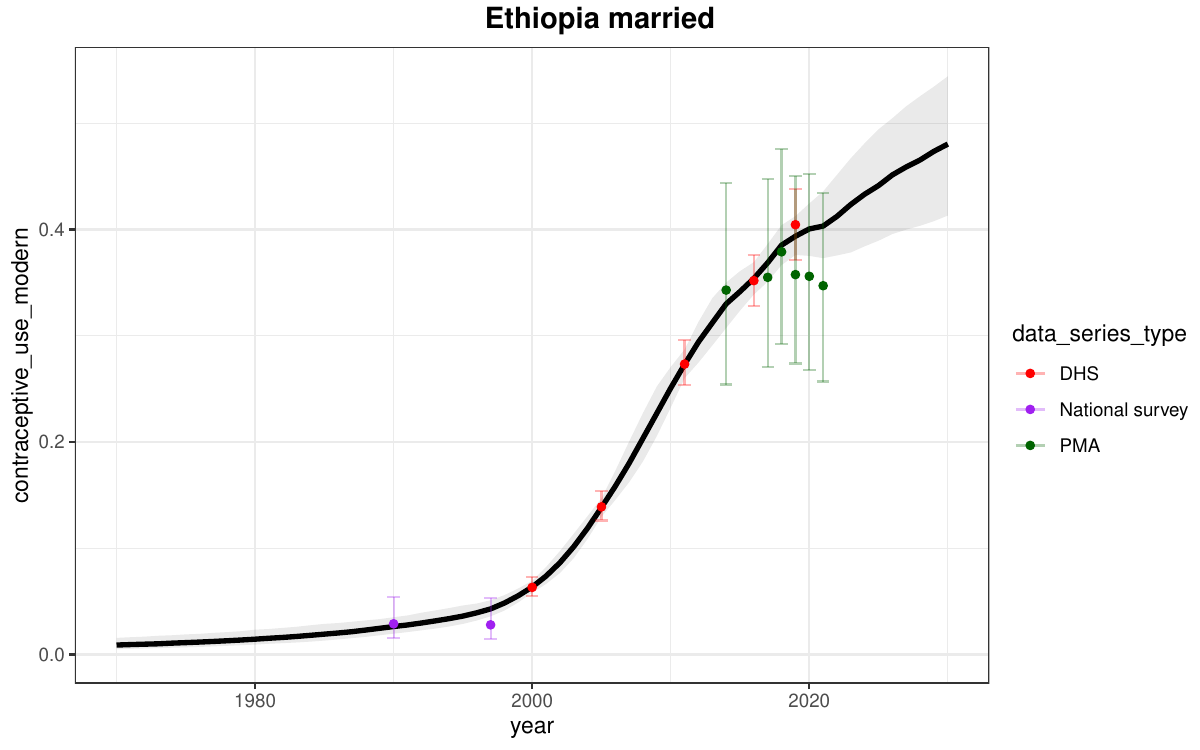}\\
     \includegraphics[width=0.45\textwidth]{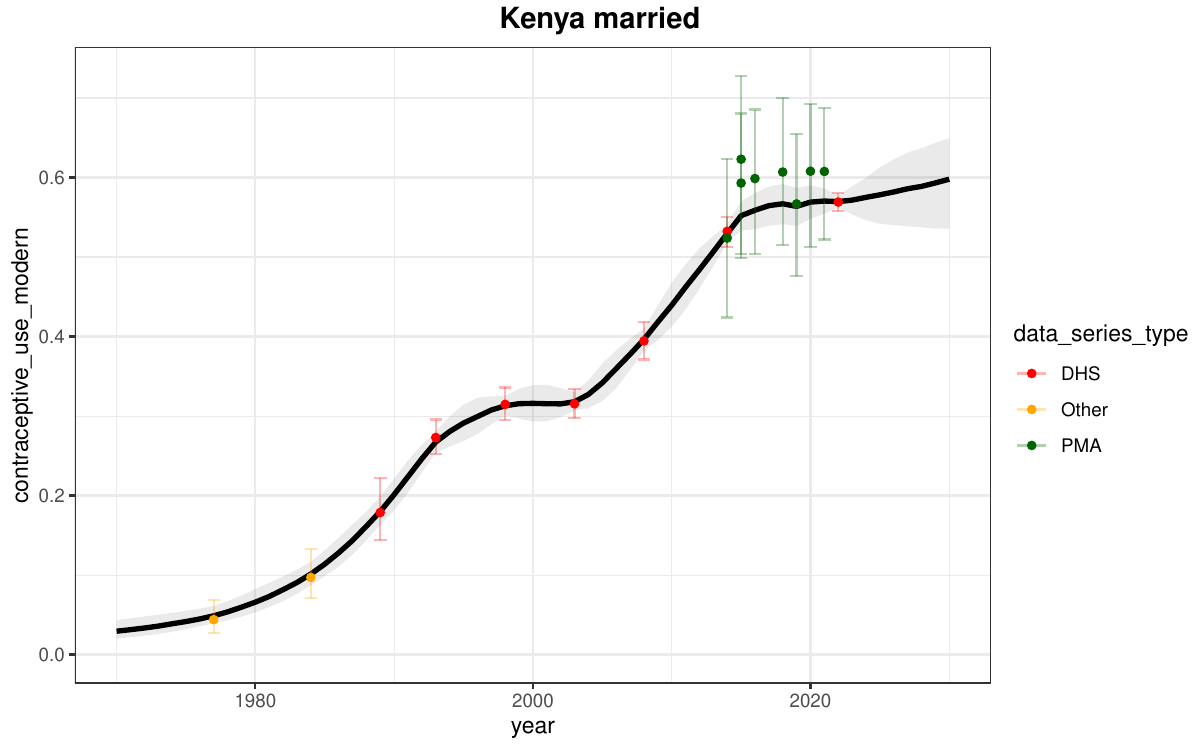} 
     \includegraphics[width=0.45\textwidth]{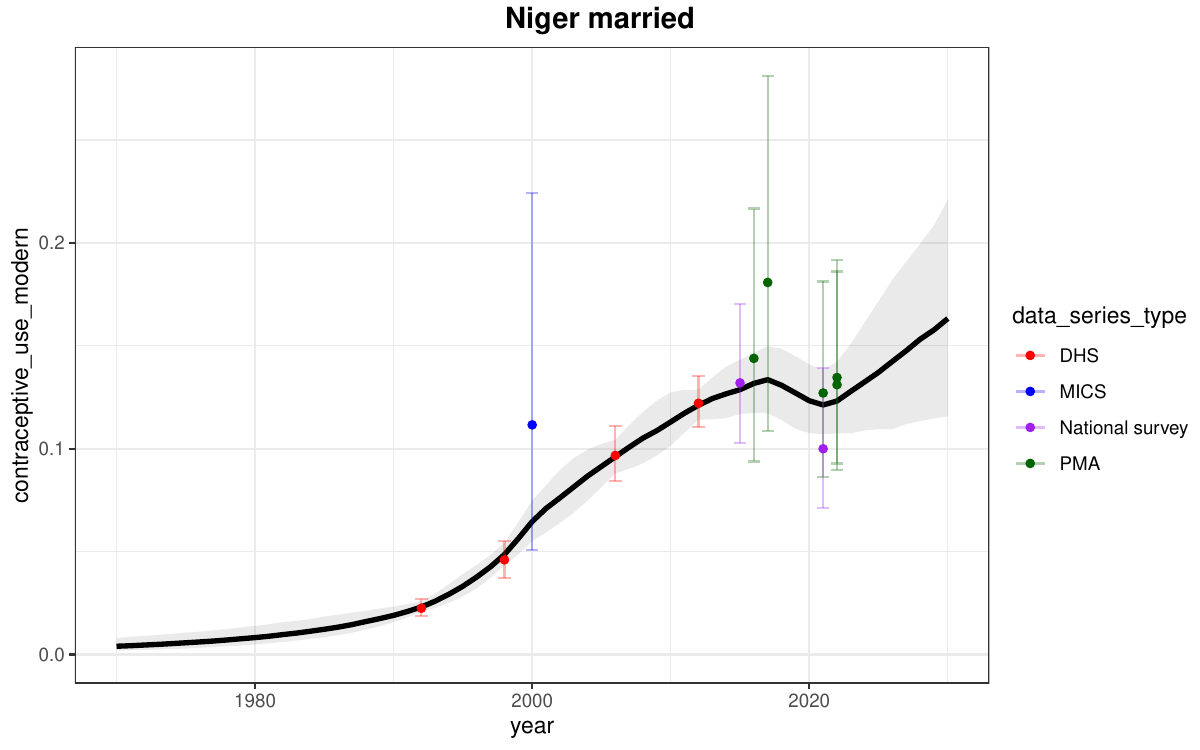} 
   \caption{\textbf{Data and estimates for mCPR among married women for Cote d'Ivoire, Ethiopia, Kenya, and Niger.} Data are shown in colored dots with 95\% confidence intervals that reflect total error variance. FPET estimates are shown in black, with grey shaded areas representing 90\% credible intervals. In the selected countries, data are available from the PMA survey program.}
    \label{fig-fits-pma}
\end{figure}

Figure~\ref{fig-fits-pma} illustrates fits for countries with PMA data. In Cote d'Ivoire, Ethiopia, and Kenya, recent DHS data suggests that the PMA series trend too high or too low.  
In Ethiopia,  the DHS data point is higher than the PMA data series for the corresponding years. FPET estimates align with DHS data and trend higher than the PMA data series. We see similar fits for the other countries with PMA data, where FPET estimates are aligned with DHS data and account for correlation in PMA series. Niger is a country with recent data from PMA and national surveys but no DHS. We see that the resulting estimates suggest some uncertainty in recent years in Niger given conflicting data from PMA and a national survey. We note that in this instance users may choose to label a more recent survey as non-outlying to produce estimates based on that type of input, for example if there is information on data quality. 

\begin{figure}[htbp]
    \centering
    \includegraphics[width=0.45\textwidth]{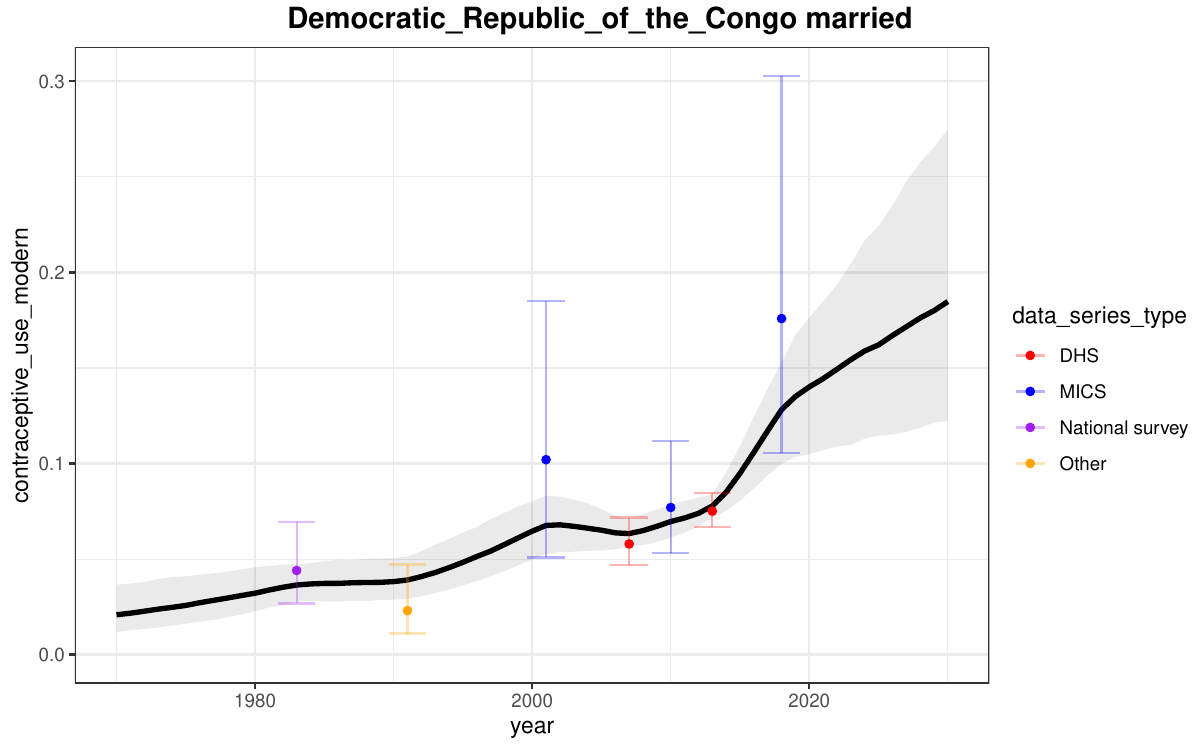} 
    \includegraphics[width=0.45\textwidth]{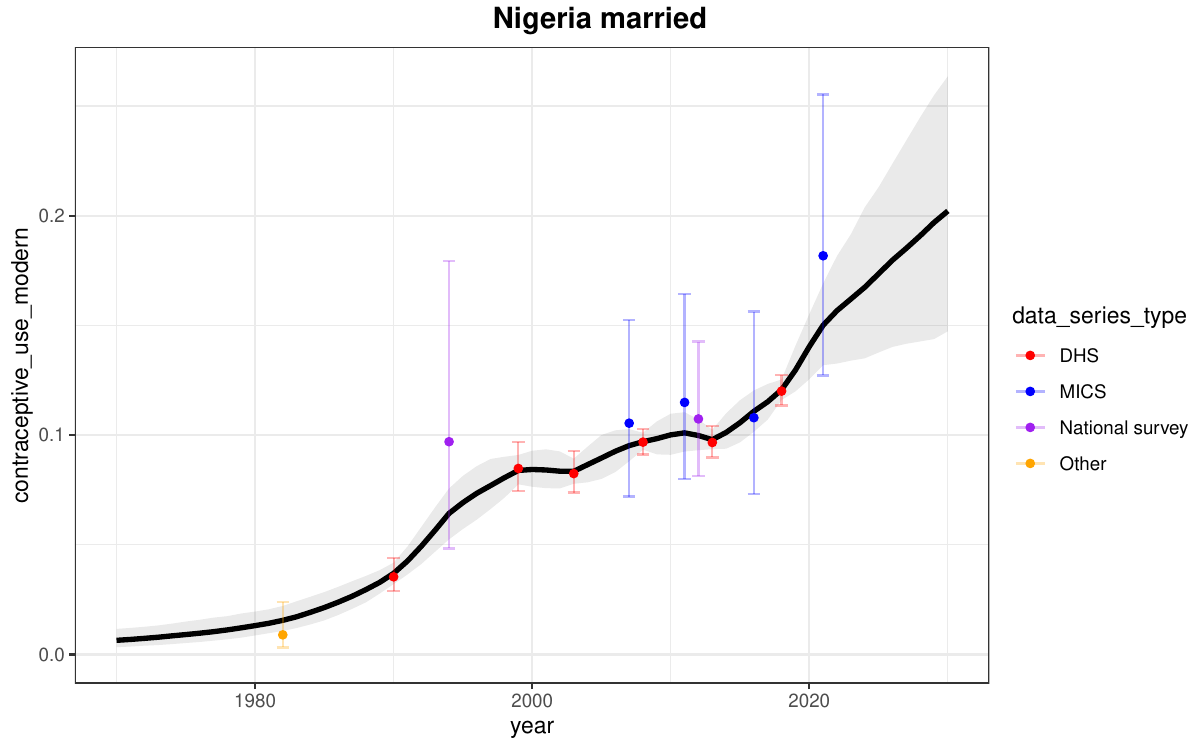}\\
     \includegraphics[width=0.45\textwidth]{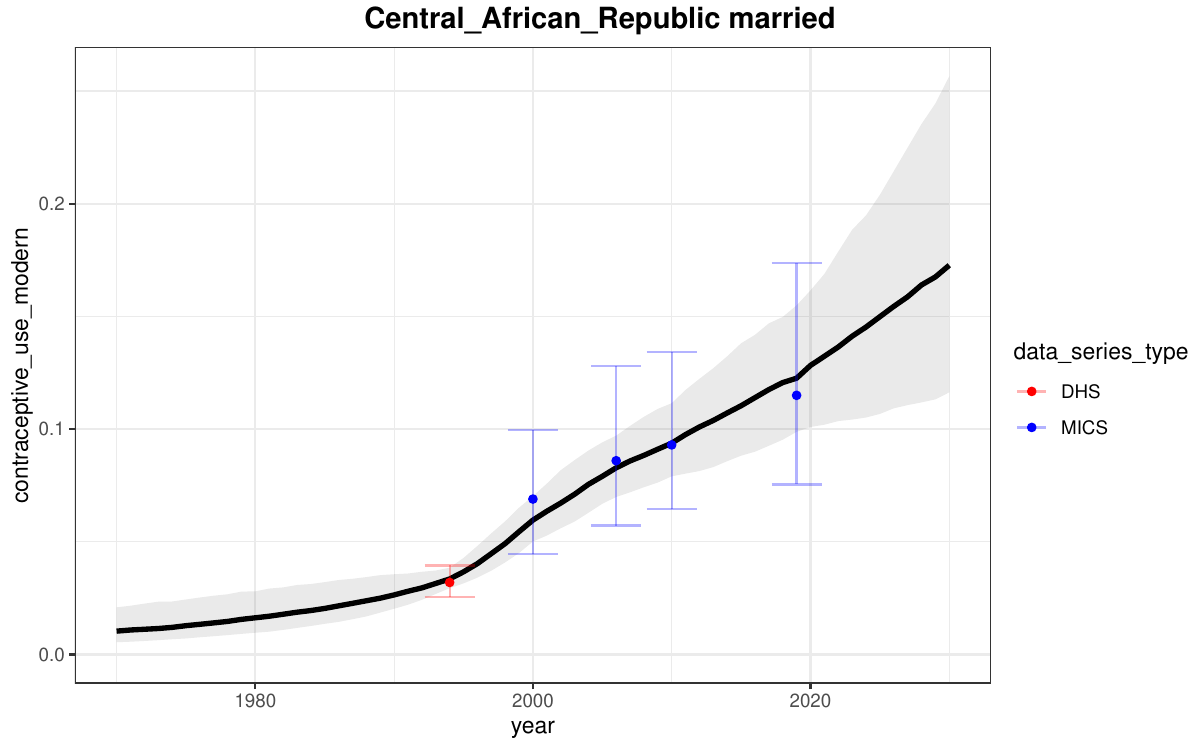} 
     \includegraphics[width=0.45\textwidth]{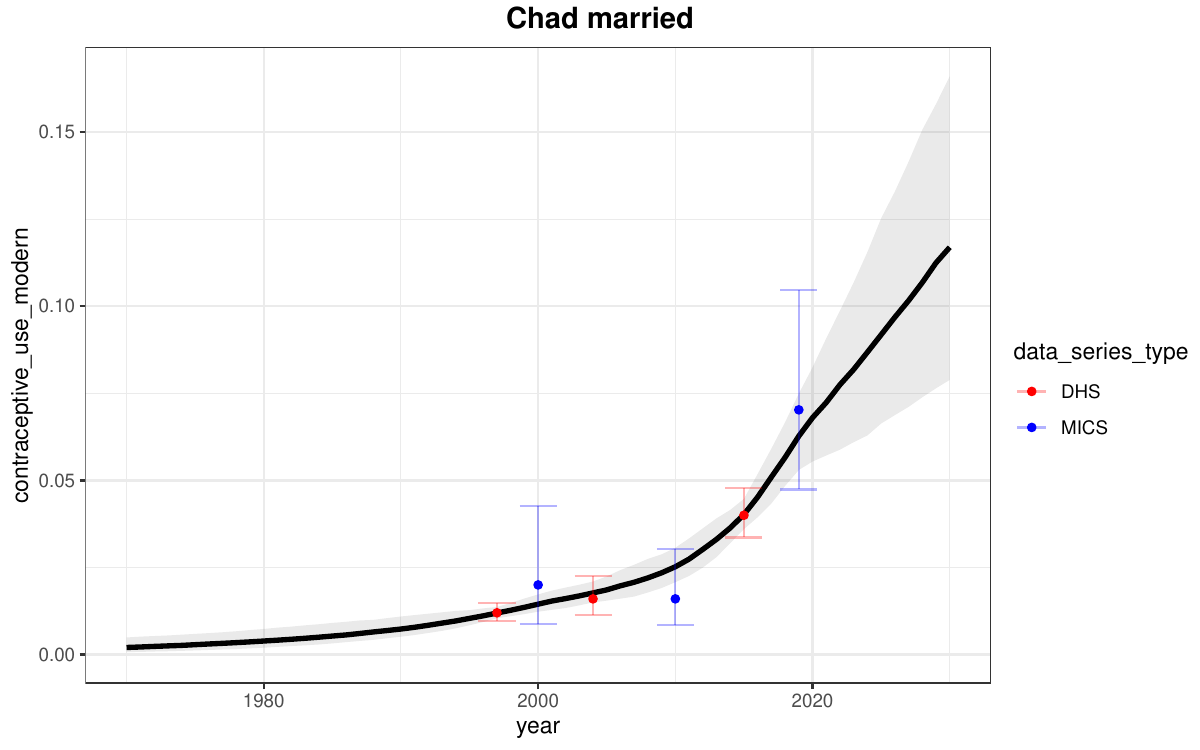} 
   \caption{\textbf{Data and estimates for mCPR among married women for Democratic Republic of the Congo, Nigeria, Central African Republic, and Chad.} Data are shown in colored dots with 95\% confidence intervals that reflect total error variance. FPET estimates are shown in black, with grey shaded areas representing 90\% credible intervals. In the selected countries, the most recent data points are from the MICS survey program. These data points have considerable uncertainty associated with them.}
    \label{fig-fits-mics}
\end{figure}

Democratic Republic of the Congo (DRC), Nigeria, Central African Republic (CAR), and Chad are examples of countries in which the most recent survey is a MICS, with large total error variance (Figure~\ref{fig-fits-mics}). In all countries, recent estimates are subject to high uncertainty. In DRC and Nigeria, the FPET estimates are lower than the MICS, suggesting that the rate of change from the most recent DHS to the most recent MICS is higher than expected. 

\section{Discussion}

In this paper we introduced a class of data models, referred to as normal-with-optional-shrinkage (NOS) models, that can be used to integrate data from multiple sources with various reporting issues. The NOS class accounts for sampling errors and differences in observational uncertainty based on survey characteristics. In addition, the data model employs horseshoe shrinkage priors to produce estimates that are robust to outlying observations. We illustrated the use of a NOS data model for the estimation of modern contraceptive use and other family planning indicators at the national level for countries using survey data. The data model estimates identified differences in error variances across survey programs, and identified outlying data points. 

The NOS data model class fits within the framework of Temporal Models for Multiple Populations (TMMPs). In this framework, an explicit distinction is made between the process model, defining the expected evolution of some indicator over time, and the data model, defining the relationship between the latent true value of the indicator and the observations. In typical applications of the TMMP class, much attention is devoted to the process model, given that it is what drives forecasts beyond the most recent observation \citep{susmann2021temporal}. However, in settings with data from multiple sources that may be subject to varying data quality, the data model is at least as important in obtaining reliable estimates of the current value of the indicator. We hope that the specification of specific data model classes, such as the NOS, will help to introduce a broader range of data models for use and evaluation. With this goal in mind, we call for the specification of additional data models to deal with other types of data and data quality issues. For example, for demographic applications, a data model class for count data from registration systems could be considered based on existing work on timing of stillbirths and maternal mortality estimation (see \cite{chong_estimating_2024, peterson_bayesian_2024}). Ultimately, we hope that expanded specification and communication of data model classes will help facilitate interpretation of existing models and  develop new ones. 

\bibliography{bibliography}

\begin{thebibliography}{12}
\providecommand{\natexlab}[1]{#1}
\providecommand{\url}[1]{\texttt{#1}}
\expandafter\ifx\csname urlstyle\endcsname\relax
  \providecommand{\doi}[1]{doi: #1}\else
  \providecommand{\doi}{doi: \begingroup \urlstyle{rm}\Url}\fi

\bibitem[Alkema et~al.(2013)Alkema, Kantorova, Menozzi, and Biddlecom]{alkema2013mcpr}
L.~Alkema, V.~Kantorova, C.~Menozzi, and A.~Biddlecom.
\newblock National, regional, and global rates and trends in contraceptive prevalence and unmet need for family planning between 1990 and 2015: a systematic and comprehensive analysis.
\newblock \emph{The Lancet}, 381\penalty0 (9878):\penalty0 1642--1652, 2022/12/21 2013.
\newblock \doi{10.1016/S0140-6736(12)62204-1}.
\newblock URL \url{https://doi.org/10.1016/S0140-6736(12)62204-1}.

\bibitem[Alkema et~al.(2024)Alkema, Susmann, Ray, Mooney, Cahill, Bietsch, Jayachandran, Kagimu, Emmart, , Mujani, Muhammad, Rosenberg, Stover, and Sonneveldt]{alkema_2024evol}
L.~Alkema, H.~Susmann, E.~Ray, S.~Mooney, N.~Cahill, K.~Bietsch, A.~Jayachandran, R.~Kagimu, P.~Emmart, , Z.~Mujani, K.~Muhammad, R.~Rosenberg, J.~Stover, and E.~Sonneveldt.
\newblock {Statistical Demography Meets Ministry of Health: The Case of the Family Planning Estimation Tool}.
\newblock 2024.

\bibitem[Cahill et~al.(2018)Cahill, Sonneveldt, Stover, Weinberger, Williamson, Wei, Brown, and Alkema]{cahill2018mcpr}
N.~Cahill, E.~Sonneveldt, J.~Stover, M.~Weinberger, J.~Williamson, C.~Wei, W.~Brown, and L.~Alkema.
\newblock Modern contraceptive use, unmet need, and demand satisfied among women of reproductive age who are married or in a union in the focus countries of the family planning 2020 initiative: a systematic analysis using the family planning estimation tool.
\newblock \emph{The Lancet}, 391\penalty0 (10123):\penalty0 870--882, 2022/12/21 2018.
\newblock \doi{10.1016/S0140-6736(17)33104-5}.
\newblock URL \url{https://doi.org/10.1016/S0140-6736(17)33104-5}.

\bibitem[Chong and Alexander(2024)]{chong_estimating_2024}
M.~Y.~C. Chong and M.~Alexander.
\newblock Estimating the timing of stillbirths in countries worldwide using a {Bayesian} hierarchical penalized splines regression model.
\newblock \emph{Journal of the Royal Statistical Society Series C: Applied Statistics}, 73\penalty0 (4):\penalty0 902--920, Aug. 2024.
\newblock ISSN 0035-9254.
\newblock \doi{10.1093/jrsssc/qlae017}.
\newblock URL \url{https://doi.org/10.1093/jrsssc/qlae017}.

\bibitem[Gabry et~al.(2024)Gabry, Češnovar, Johnson, and Bronder]{cmdstanr2022}
J.~Gabry, R.~Češnovar, A.~Johnson, and S.~Bronder.
\newblock \emph{cmdstanr: R Interface to 'CmdStan'}, 2024.
\newblock URL \url{https://mc-stan.org/cmdstanr/}.
\newblock R package version 0.8.1, https://discourse.mc-stan.org.

\bibitem[Kantorová et~al.(2020)Kantorová, Wheldon, Ueffing, and Dasgupta]{kantorova_estimating_2020}
V.~Kantorová, M.~C. Wheldon, P.~Ueffing, and A.~N.~Z. Dasgupta.
\newblock Estimating progress towards meeting women’s contraceptive needs in 185 countries: {A} {Bayesian} hierarchical modelling study.
\newblock \emph{PLOS Medicine}, 17\penalty0 (2):\penalty0 e1003026, Feb. 2020.
\newblock ISSN 1549-1676.
\newblock \doi{10.1371/journal.pmed.1003026}.
\newblock URL \url{https://journals.plos.org/plosmedicine/article?id=10.1371/journal.pmed.1003026}.
\newblock Publisher: Public Library of Science.

\bibitem[New et~al.(2017)New, Cahill, Stover, Gupta, and Alkema]{new_levels_2017}
J.~R. New, N.~Cahill, J.~Stover, Y.~P. Gupta, and L.~Alkema.
\newblock Levels and trends in contraceptive prevalence, unmet need, and demand for family planning for 29 states and union territories in {India}: a modelling study using the {Family} {Planning} {Estimation} {Tool}.
\newblock \emph{The Lancet Global Health}, 5\penalty0 (3):\penalty0 e350--e358, Mar. 2017.
\newblock ISSN 2214-109X.
\newblock \doi{10.1016/S2214-109X(17)30033-5}.
\newblock URL \url{http://www.sciencedirect.com/science/article/pii/S2214109X17300335}.

\bibitem[Peterson et~al.(2024)Peterson, Guranich, Cresswell, and Alkema]{peterson_bayesian_2024}
E.~N. Peterson, G.~Guranich, J.~A. Cresswell, and L.~Alkema.
\newblock A {Bayesian} {Approach} to {Estimate} {Maternal} {Mortality} {Globally} {Using} {National} {Civil} {Registration} {Vital} {Statistics} {Data} {Accounting} for {Reporting} {Errors}.
\newblock \emph{Statistics and Public Policy}, 11\penalty0 (1):\penalty0 2286313, Dec. 2024.
\newblock \doi{10.1080/2330443X.2023.2286313}.
\newblock URL \url{https://doi.org/10.1080/2330443X.2023.2286313}.

\bibitem[Piironen and Vehtari(2017)]{piironen2017horseshoe}
J.~Piironen and A.~Vehtari.
\newblock {Sparsity information and regularization in the horseshoe and other shrinkage priors}.
\newblock \emph{Electronic Journal of Statistics}, 11\penalty0 (2):\penalty0 5018 -- 5051, 2017.
\newblock \doi{10.1214/17-EJS1337SI}.
\newblock URL \url{https://doi.org/10.1214/17-EJS1337SI}.

\bibitem[{R Core Team}(2022)]{r2022}
{R Core Team}.
\newblock \emph{R: A Language and Environment for Statistical Computing}.
\newblock R Foundation for Statistical Computing, Vienna, Austria, 2022.
\newblock URL \url{https://www.R-project.org/}.

\bibitem[{Stan Development Team}(2024)]{stan2023}
{Stan Development Team}.
\newblock Stan modeling language users guide and reference manual, 2.35, 2024.
\newblock URL \url{https://mc-stan.org}.

\bibitem[Susmann et~al.(2022)Susmann, Alexander, and Alkema]{susmann2021temporal}
H.~Susmann, M.~Alexander, and L.~Alkema.
\newblock Temporal models for demographic and global health outcomes in multiple populations: Introducing a new framework to review and standardise documentation of model assumptions and facilitate model comparison.
\newblock 2022.
\newblock \doi{10.1111/insr.12491}.

\end{thebibliography}

\section{Appendix}

\subsection{Identifying possible outliers}
The algorithm to classify whether observations are possibly outlying includes three steps. 

For each country, we first label observations as possibly outlying if there are data quality concerns or other documented substantive information to suggest they may be subject to substantial error. For example, DHS observations collected before 1990 may be outlying because of possible misclassification of women who were pregnant or post-partum. A survey may be classified as being possible outlying if issues with the survey process have been documented. 

Subsequently, for each country, we check if we can identify a reference source category among the left over points, which, as the name suggest, provides a reference in the estimation exercise and is assumed to be not subject to outlier errors. This category is DHS data, if available. If no DHS data are available, data from national surveys or other survey programs or used, with the choice based on the source type with the most observations after 1990 in that country. 
If the algorithm does not result in a reference category, i.e., if there are no non-possibly-outlying observations that are from DHS, national, or other surveys, then the population does not have a reference category and all observations are possibly outlying. 

Finally, we use 
estimates of longer term trends to identify additional points that were initially classified into the reference category but may be outlying due to measurement issues. The rationale for this final step in categorization is as follows: For FP indicators, extreme fluctuations are more likely to be data errors. Hence we flag observations that are extreme as compared to smooth long-term trends as they may be subject to data quality issues, and allow these observations to be possibly outlying in final model fitting. To flag observations, we construct estimates of long term trends (see \citep{alkema_2024evol}). We then calculate the total error relative to longer term trends as the difference between the transformed observation and  the long-term trend estimate. We label those observations with absolute errors among the 10\% of largest absolute errors as possibly outlying.

\subsection{Results for all countries}

\begin{figure}[h]   \centering
   \caption{\textbf{Data and estimates for mCPR among married women for all countries in the FP2030 initiative.} Data are shown in colored dots with 95\% confidence intervals that reflect total error variance. FPET estimates are shown in black, with grey shaded areas representing 90\% credible intervals.}\label{fig-fits}
\end{figure}

 \clearpage

 \includepdf[pages=1-,nup=2x3,pagecommand=,width=0.46\columnwidth]{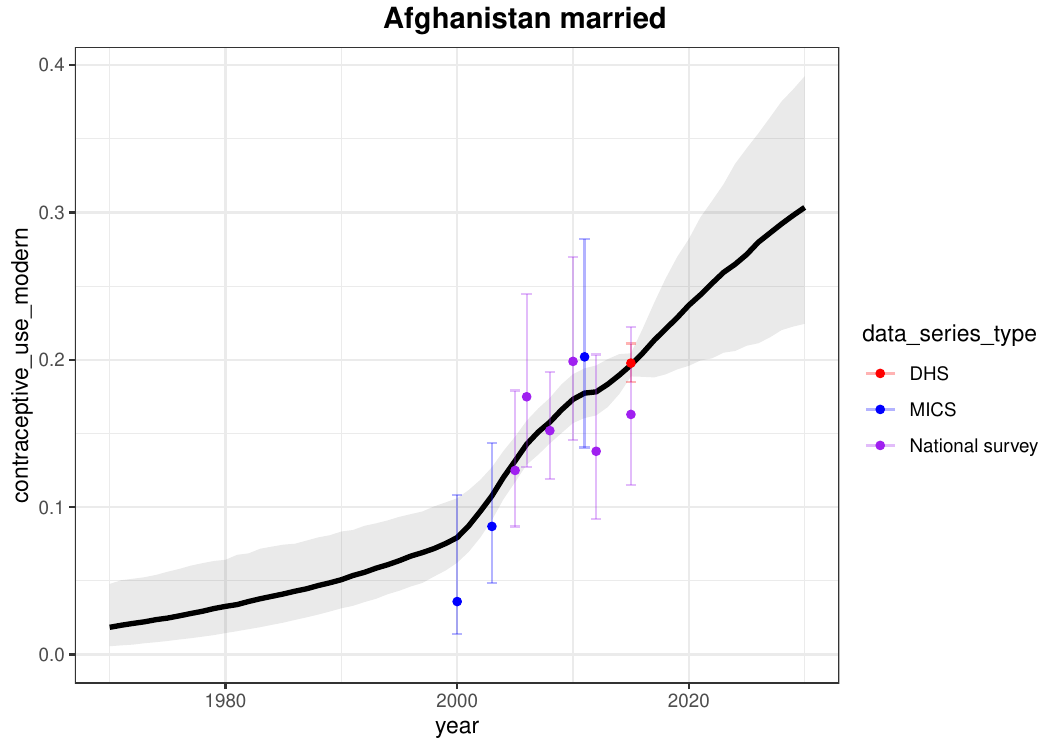}

\end{document}